           \documentstyle[prl,preprint,aps]{revtex}

\begin{document}

\title{
                     The Updated MSW Analysis and \\
		     the Standard Solar Model Uncertainties
\footnote{%
                               This is an update of a talk presented at
                               SUSY93, Northeastern University, Boston, MA,
                               April 1993.
}
}
\author{
                     Naoya Hata and Paul Langacker
\\
}
\address{                                                         
                     Department of Physics,                       
                     University of Pennsylvania,                  
                     Philadelphia, PA 19104                       
}                                                                 
\date{
                     August 6, 1993, UPR-0581T 
}
\maketitle


\begin{abstract}

We update the analysis of the MSW  and general astrophysical solutions
to the combined solar neutrino observations by including the
GALLEX II result.   We also show that our parametrized flux uncertainties
are equivalent to the Monte-Carlo results of Bahcall and Ulrich.

\end{abstract}

\newpage


\section{Global Analysis Updated}

\subsection{MSW}

By including the GALLEX II data \cite{GALLEX-II}, we update our previous MSW
analysis \cite{BHKL,HL}
which includes a joint $\chi^2$ analysis of all experiments, a full
treatment of theoretical uncertainties including correlations between
experiments, and the Earth effect. (See also \cite
{Bahcall-Haxton,Shi-Schramm,Gelb-Kwong-Rosen,Krastev-Petcov,KGW}.)  The
experimental data used in the analysis are summarized in
Table \ref{tab_expdata}.  The combined result of the Homestake
\cite{Homestake}, gallium \cite{SAGE,GALLEX,GALLEX-II},
and Kamiokande  (including the day-night data) \cite{Kamiokande} experiments
are shown in Fig.~\ref{fig_daynight}.  The allowed large-angle region for
$\Delta m^2 \sim 10^{-5} \mbox{eV}^2$ has been reduced significantly by
including the new GALLEX result.  The allowed regions
\footnote{
We define the 90, 95, and 99\% C.L. regions by increases of 4.6, 6.0, and
9.2 in the overall $\chi^2$ with respect to the global minimum.  This
prescription is only rigorously valid when the probability distribution in
the parameter space is gaussian around the global minimum.  Improved
treatments will be discussed in a subsequent paper \cite{HL-uncertainties}.
}
at 90, 95, and 99\% C.L. are shown in Fig.~\ref{fig_CLs};
there is a third allowed solution in the large-angle region at 99\% C.L for
$\Delta m^2 \sim 10^{-7} \mbox{eV}^2$.
In Fig.~\ref{fig_earth} we show the combined result including the Earth
effect, but without the Kamiokande day-night data.

\subsection{General Astrophysical Solution}

When the solar neutrino data are fit to a general solar model \cite{BHL},
using the $pp$, CNO, $^7$Be, and $^8$B fluxes as free parameters with
the luminosity constraint imposed, the best fit value requires
$\phi(\mbox{Be}) / \phi(\mbox{Be})_{SSM} < 0.08$ and
$\phi(\mbox{B})  / \phi(\mbox{B})_{SSM} = 0.37 \pm 0.04$, but with a large
$\chi^2$ ( = 5.1 for 1 degrees of freedom); this solution is
excluded at 98\% C.L.

\section{The parametrized SSM Uncertainties}

In the MSW analysis it is important to include the Standard Solar Model (SSM)
flux uncertainties
properly, since their magnitudes are comparable to the experimental
uncertainties and also strongly correlated from experiment to experiment.
One way to incorporate those SSM uncertainties is to carry out Monte
Carlo simulations, as done by Bahcall and Haxton \cite{Bahcall-Haxton}, which
is robust but time
consuming.  Alternatively, we parameterize the
flux uncertainties with the uncertainties of the central temperature and of the
the relevant nuclear reaction cross sections.  This method was, however,
questioned  by Bahcall on the grounds  that this simplification might fail in
estimating the uncertainties accurately \cite{Bahcall-Texas}.  We have
numerically compared the two methods
and conclude that they yields almost identical results \cite{HL-uncertainties}.

First we have compared the uncertainties of the major fluxes ($pp$, $^7$Be,
and $^8$B) and their correlations.  The results are listed in
Table~\ref{tab_magnitudes} for the magnitudes and Table~\ref{tab_correlations}
for the correlations.  The results are also graphically shown in
Fig.~\ref{fig_BeB}.  The numerical results show that the parameterization
method accurately reproduces the Monte-Carlo flux uncertainties.

We have also carried out the comparison for the uncertainties of the rate
predictions for the experiments.  The magnitudes and the correlations of the
uncertainties are compared in Table~\ref{tab_exp_magnitudes} and
Table~\ref{tab_exp_correlations}, respectively.  Again the agreement is
excellent.

\newpage



\begin{table}[p]
\caption{
%
%
The standard solar model predictions of Bahcall and Pinsonneault
\protect\cite{Bahcall-Pinsonneault} and of Turck-Chi\'eze and Lopes
\protect\cite{Turck-Chieze-Lopes}, along with the results of the solar
neutrino experiments.  The gallium experiment is the combined result of
SAGE ($58 ^{+22}_{-27} \, \mbox{SNU}$) and GALLEX I and II
($87 \pm 16 \, \mbox{SNU}$).
}
\label{tab_expdata}
\vspace{1.0ex}
\begin{tabular}{l  c c c}
%
               & BP SSM        & TCL SSM      & Experiments \\
\hline
Kamiokande     &  1 $\pm$ 0.14 & 0.77$\pm$0.19   & 0.50$\pm$0.07 BP-SSM  \\
Homestake (Cl) &  8$\pm$1 SNU  & 6.4$\pm$1.4 SNU & 2.23$\pm$0.23 SNU
                                                  (0.28$\pm$0.03 BP-SSM) \\
SAGE \& GALLEX (Ga) & 131.5$^{+7}_{-6}$ SNU & 122.5$\pm$7 SNU & 77$\pm$13 SNU
                                                      (0.59$\pm$0.10 BP-SSM) \\
%
\end{tabular}
\end{table}

\vspace{5ex}
\begin{table}[p]
\caption{
%
%
The magnitudes of flux uncertainties ($\Delta \phi / \phi$ at $1 \sigma$)
of the Bahcall-Ulrich SSM \protect\cite{Bahcall-Ulrich}, Bahcall-Ulrich
Monte Carlo SSMs, and the parametrized SSM using the central temperature and
the nuclear reaction cross sections.  Also listed are the
uncertainties of the Bahcall-Pinsonneault SSM and of its parametrized fluxes.
(The Monte Carlo study of the Bahcall-Pinsonneault model is not available.)
}
\label{tab_magnitudes}
\vspace{1.0ex}
\begin{tabular}{ l  c c c c }
%
                                     &    $pp$     &   $^7$Be     &  $^8$B   \\
\hline
Bahcall-Ulrich SSM                   &  0.0059     &   0.050      &  0.12   \\
Bahcall-Ulrich SSM (Monte Carlo)     &  0.0067     &   0.05       &  0.17   \\
Parametrized ($\Delta T_C = 0.0053$) &  0.0068     &   0.05       &  0.12   \\
\hline
Bahcall-Pinsonneault SSM             &  0.0067     &   0.06       &  0.14   \\
Parametrized ($\Delta T_C = 0.0057$) &  0.007      &   0.06       &  0.14   \\
%
\end{tabular}
\end{table}

\begin{table}[p]
\caption{
%
%
The correlation matrices of flux uncertainties obtained from the Bahcall-Ulrich
Monte Carlo SSMs and of the parameterization method.  The agreement between
the two methods is
excellent except the $pp$ -- $^7$Be element.  This can be improved by
anticorrelating the uncertainties from the $S_{11}$ for the two fluxes,
which is a consequence of the luminosity constraint.  The effect of
anticorrelating $S_{11}$ is completely negligible in the MSW fits, and we
present the MSW results without the anticorrelation.
}
\label{tab_correlations}
\vspace{1.0ex}
\begin{tabular}{ l  c c c c }
%
                                     &    $pp$     &  $^8$B       &  $^7$Be  \\
\hline
\multicolumn{1}{l}{Bahcall-Ulrich SSM (Monte Carlo)}     &&                \\
$pp$                                 &   1         &              &         \\
$^8$B                                &$-$0.726     &   1          &         \\
$^7$Be                               &$-$0.917     &   0.742      &   1     \\
\hline
\multicolumn{3}{l}{Parametrized with $\Delta T_C$ and $\Delta s$}           \\
$pp$                                 &   1         &              &         \\
$^8$B                                &$-$0.679     &   1          &         \\
$^7$Be                               &$-$0.677     &   0.773      &   1     \\
\hline
\multicolumn{3}{l}{Parametrized with $\Delta T_C$ and $\Delta s$ ( $S_{11}$
is anticorrelated between $pp$ and $^7$Be)}      \\
$pp$                                 &   1         &              &         \\
$^8$B                                &$-$0.679     &   1          &         \\
$^7$Be                               &$-$0.916     &   0.773      &   1     \\
%
\end{tabular}
\end{table}

\begin{table}[p]
\caption{
%
%
The comparison of the magnitudes of rate uncertainties for the Bahcall-Ulrich
SSM, Monte Carlo SSMs, and the parametrized SSM.
Also listed are the uncertainties in the Bahcall-Pinsonneault SSM and its
parametrized SSM.
}
\label{tab_exp_magnitudes}
\vspace{1.0ex}
\begin{tabular}{ l  c c c c }
%
                                     &  Kamiokande &      Cl      &  Ga   \\
\hline
Bahcall-Ulrich SSM                   &  0.12       &  0.11     &$+0.05 -0.04$\\
Monte Carlo                          &  0.12       &  0.11        &   0.05  \\
Parametrized ($\Delta T_C = 0.0053$) &  0.12       &  0.12        &   0.05  \\
\hline
Bahcall-Pinsonneault SSM             &  0.14       &   0.13       &  0.05   \\
Parametrized ($\Delta T_C = 0.0057$) &  0.14       &   0.13       &  0.05   \\
%
\end{tabular}
\end{table}

\newpage
\begin{table}[p]
\caption{
%
%
The comparison of the correlations of rate uncertainties for Monte Carlo SSMs
and the parametrized SSM.
Also listed is  the parametrized SSM with the $S_{11}$ uncertainty
anticorrelated between the $pp$ and $^7$Be fluxes.
}
\label{tab_exp_correlations}
\vspace{1.0ex}
\begin{tabular}{ l  c c c c }
%
                                  &    Kamiokande     &   Cl     &  Ga   \\
\hline
\multicolumn{3}{l}{Bahcall-Ulrich SSM (Monte Carlo)}                       \\
Kamiokande                        &   1         &              &         \\
Cl                                &$-$0.997     &   1          &         \\
Ga                                &$-$0.920     &   0.947      &   1     \\
\hline
\multicolumn{3}{l}{Parametrized with $\Delta T_C$ and $\Delta s$}           \\
Kamiokande                        &   1         &              &         \\
Cl                                &$-$0.995     &   1          &         \\
Ga                                &$-$0.888     &   0.928      &   1     \\
\hline
\multicolumn{3}{l}{Parametrized with $\Delta T_C$ and $\Delta s$, with $S_{11}$
anticorrelated between $pp$ and $^7$Be}      \\
Kamiokande                        &   1         &              &         \\
Cl                                &$-$0.995     &   1          &         \\
Ga                                &$-$0.898     &   0.936      &   1     \\
%
\end{tabular}
\end{table}


%
%
\begin{figure}[p]
\vspace{2ex}
\caption{
The allowed regions of the combined  Homestake, Kamiokande, and gallium
observations including the Kamiokande day-night data.  The region excluded
by the day-night data is also shown (dotted line).
}
\label{fig_daynight}
\end{figure}

%
%
\begin{figure}[p]
\vspace{2ex}
\caption{
The allowed regions of the combined fit at 90, 95, and 99\% C.L.  There is a
third  solution in the large-angle region for
$\Delta m^2 \sim 10^{-7} \mbox{eV}^2$ at 99\% C.L.
}
\label{fig_CLs}
\end{figure}

%
%
\begin{figure}[p]
\vspace{2ex}
\caption{
The allowed region with the Earth effect, but without the Kamiokande
day-night data.
}
\label{fig_earth}
\end{figure}

%
%
\begin{figure}[p]
\vspace{2ex}
\caption{
The distribution of the $pp$, $^7$Be, and $^8$B flux of the Bahcall-Ulrich
Monte Carlo SSMs (histograms) and the parametrized method (solid curves)
that assumes
Gaussian distributions of the central temperatures and the nuclear reaction
cross sections around their central values.
}
\label{fig_fluxes}
\end{figure}

%
%
\begin{figure}[p]
\vspace{2ex}
\caption{
The distributions of the expected rate for the Kamiokande, chlorine, and
gallium experiments obtained from Bahcall-Ulrich Monte Carlo fluxes
(histograms) and the parametrized fluxes (solid curve).  In both cases the
detector cross section uncertainties are not included. (They are included
in the analysis of the MSW and of general astrophysical solutions.)
}
\label{fig_rates}
\end{figure}

%
%
\begin{figure}[p]
\vspace{2ex}
\caption{
The distributions of the $^7$Be and $^8$B flux of the Bahcall-Ulrich SSMs
(dots) and the 90\% C.L. contour of the parametrized SSM (solid curve).
The magnitudes and the correlations are the same for the two methods.
}
\label{fig_BeB}
\end{figure}

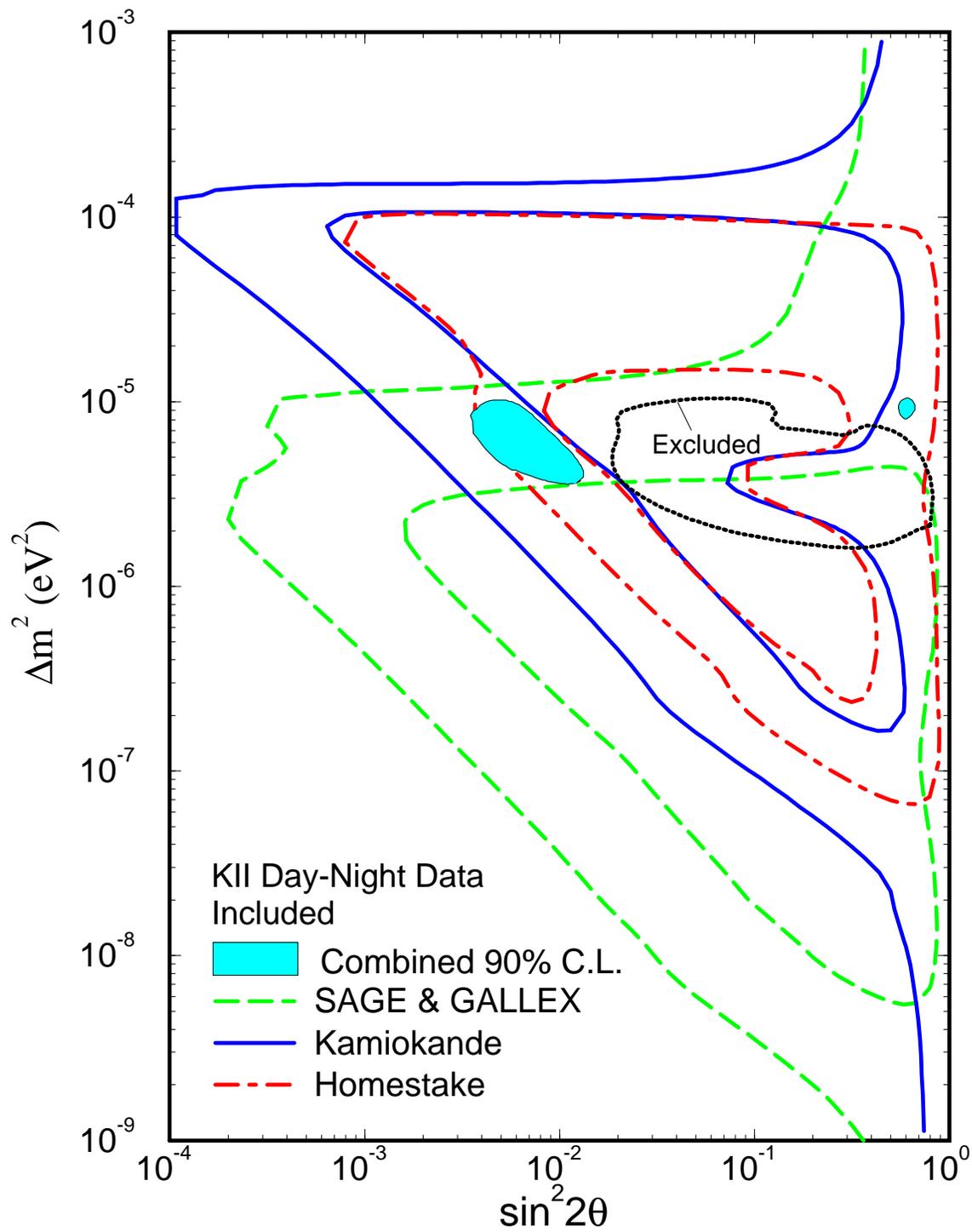
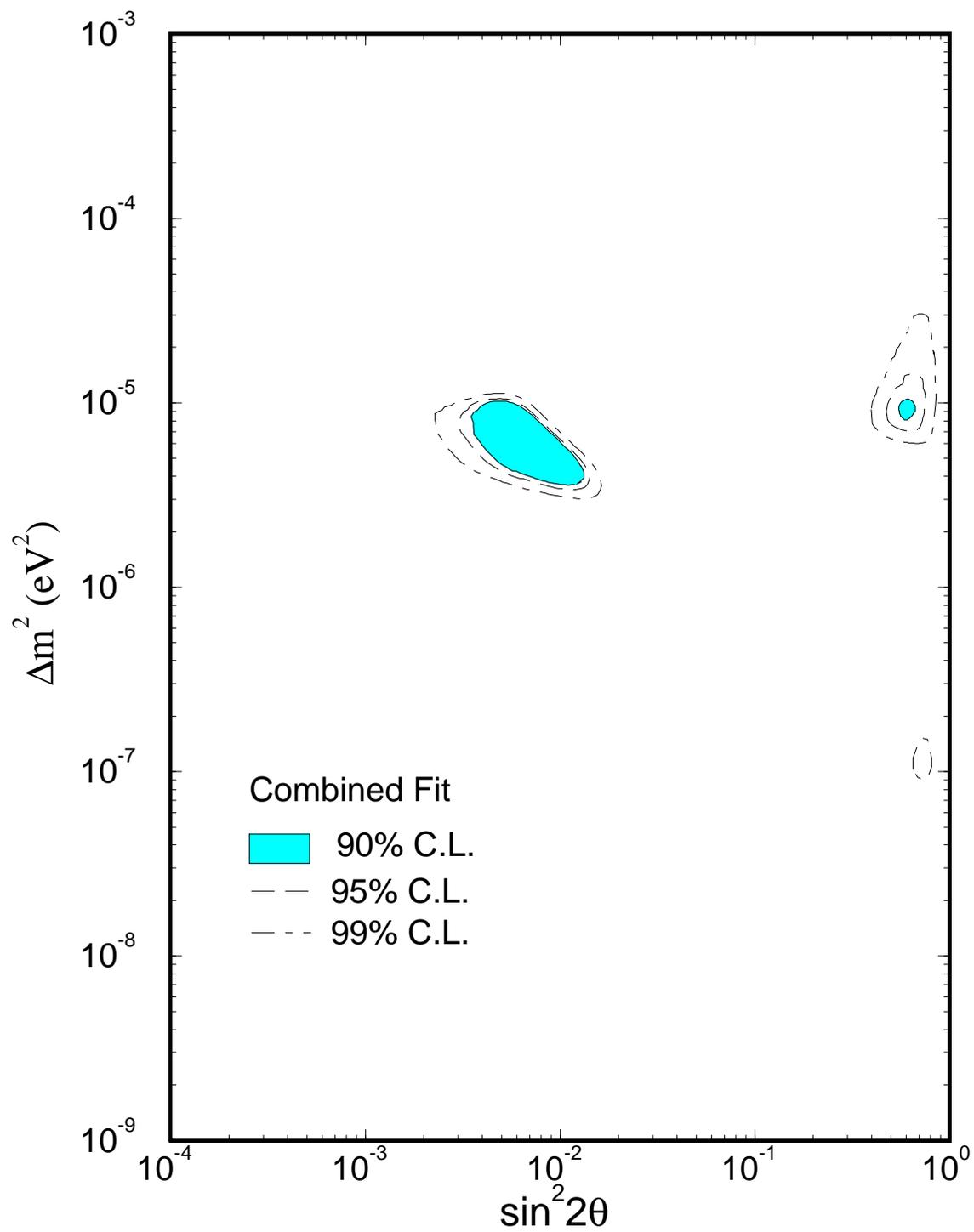
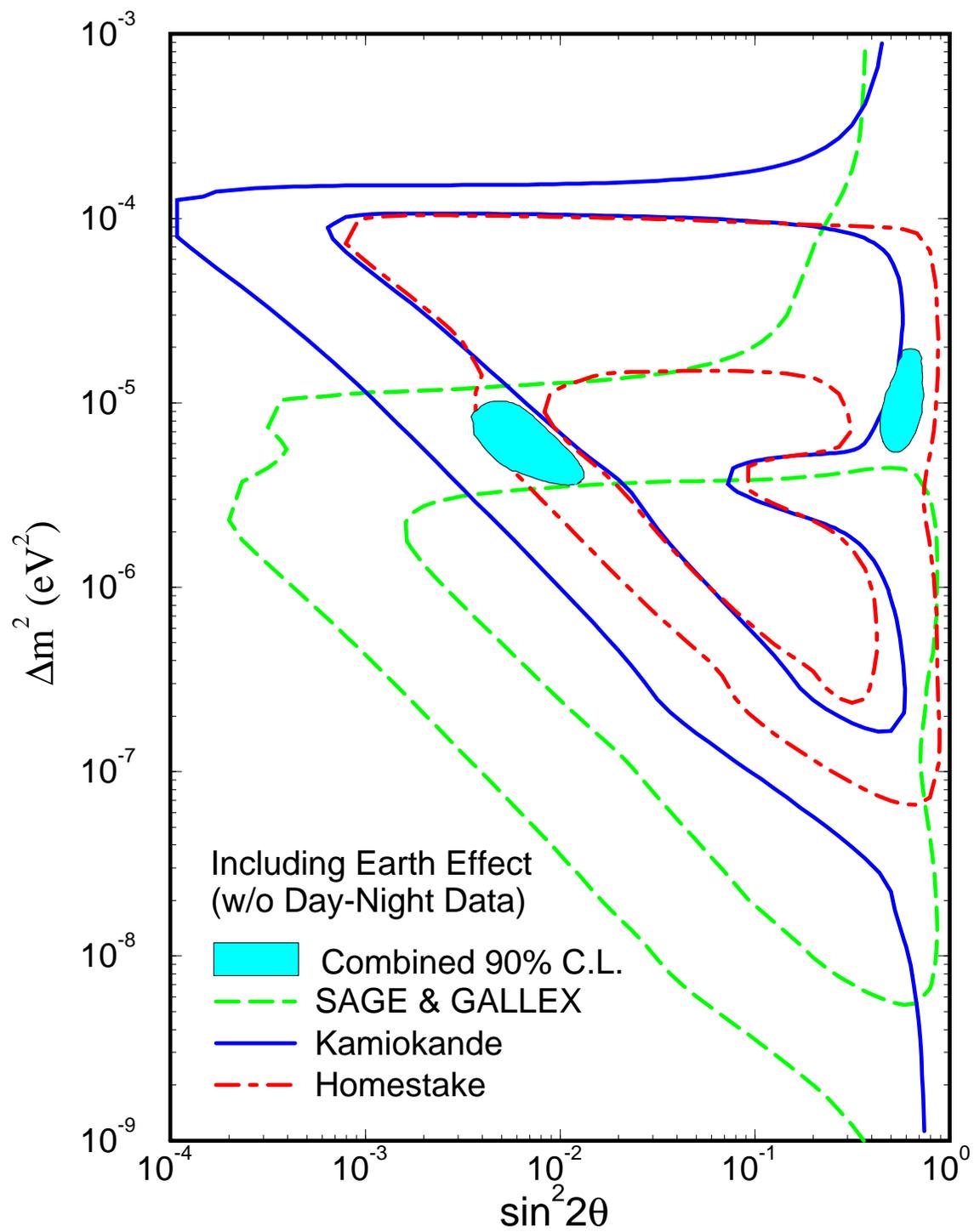
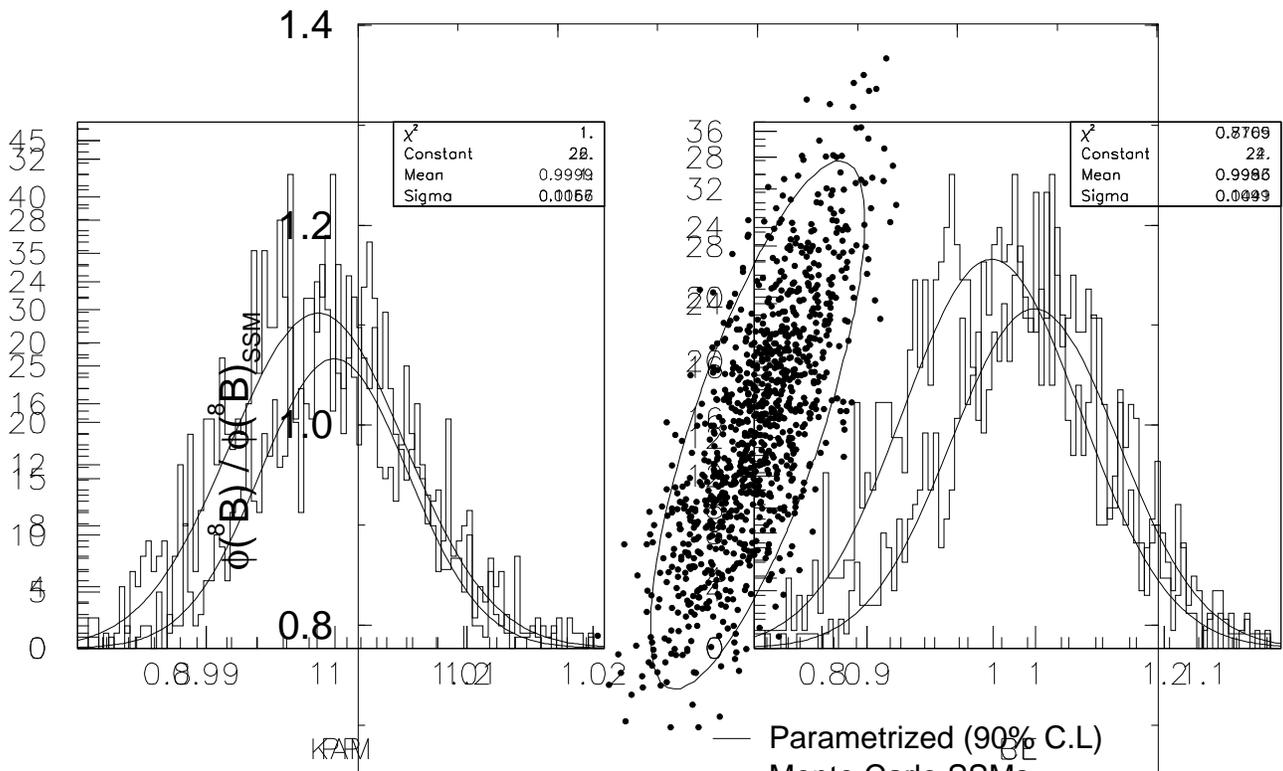
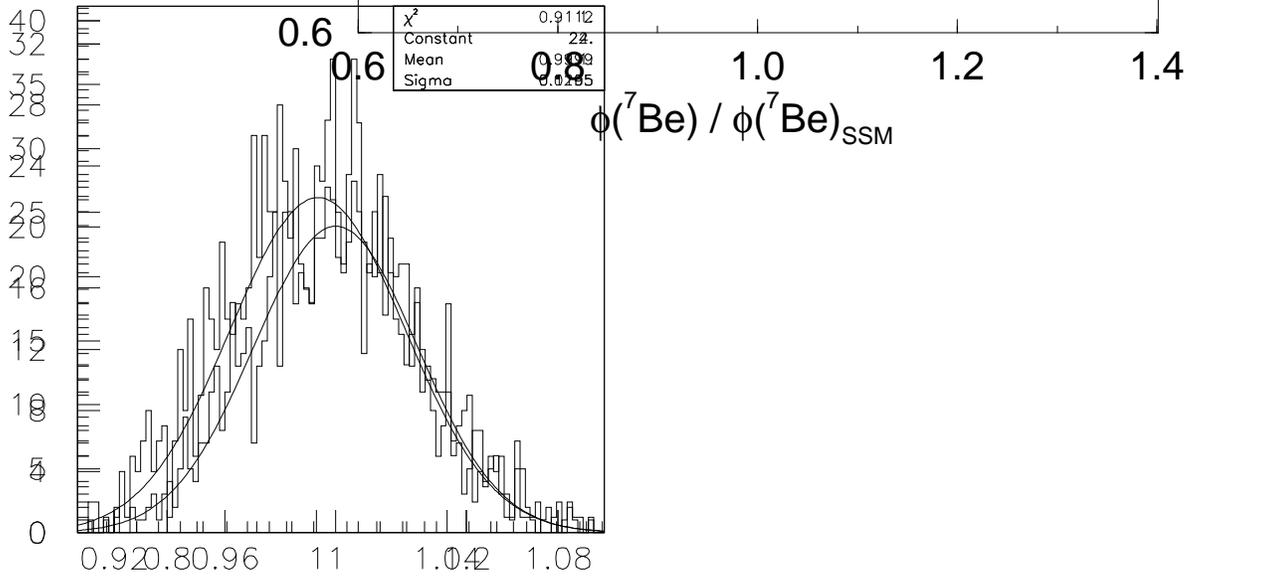
\end{document}